\documentclass[journal=jacsat,manuscript=article]{achemso}

\usepackage[version=3]{mhchem} 
\usepackage{xcolor}
\usepackage{soul}

\author{Ofir Blumer}
\affiliation[TAU]
{School of Chemistry, Tel Aviv University, Tel Aviv 6997801, Israel.}
\author{Shlomi Reuveni}
\affiliation[TAU]
{School of Chemistry, Tel Aviv University, Tel Aviv 6997801, Israel.}
\alsoaffiliation[Sackler]
{The Center for Computational Molecular and Materials
Science, Tel Aviv University, Tel Aviv 6997801, Israel.}
\alsoaffiliation[BioSoft]
{The Center for Physics and Chemistry of Living Systems, Tel Aviv University, Tel Aviv 6997801, Israel.}
\author{Barak Hirshberg}
\email{hirshb@tauex.tau.ac.il}
\affiliation[TAU]
{School of Chemistry, Tel Aviv University, Tel Aviv 6997801, Israel.}
\alsoaffiliation[Sackler]
{The Center for Computational Molecular and Materials
Science, Tel Aviv University, Tel Aviv 6997801, Israel.}

\title[]{Resetting Metadynamics}

\abbreviations{IR,NMR,UV}

\begin{document}

\begin{abstract}

Metadynamics is a powerful method to accelerate molecular dynamics simulations, but its efficiency critically depends on the identification of collective variables that capture the slow modes of the process. Unfortunately, collective variables are usually not known a priori and finding them can be very challenging. We recently presented a collective variables-free approach to enhanced sampling using stochastic resetting. Here, we combine the two methods for the first time, showing that it can lead to greater acceleration than either of them separately. We also demonstrate that resetting Metadynamics simulations performed with \textit{suboptimal} collective variables can lead to speedups comparable with those obtained with \textit{optimal} collective variables. Therefore, the application of stochastic resetting can be an alternative to the challenging task of improving suboptimal collective variables, at almost no additional computational cost. Finally, we propose a method to extract unbiased mean first-passage times from Metadynamics simulations with resetting, resulting in improved tradeoff between speedup and accuracy. This work opens the way for combining stochastic resetting with other enhanced sampling methods to accelerate a broad range of molecular simulations. 

\end{abstract}

\section{Introduction}

Molecular dynamics (MD) simulations provide valuable insights into
the dynamics of complex chemical and physical systems. 
They are a powerful tool but, due to their atomic spatial and temporal resolution, they
 cannot be applied to processes that are longer than a few microseconds, 
 such as protein folding and crystal nucleation\cite{salvalaglio_assessing_2014,tiwary_metadynamics_2013}. 
 Different methods have been developed in order to overcome this timescale problem,
 such as umbrella sampling\cite{torrie_nonphysical_1977,kastner_umbrella_2011}, replica-exchange\cite{sugita_replica-exchange_1999}, free energy dynamics\cite{rosso_adiabatic_2002,rosso_use_2002}, Metadynamics (MetaD)\cite{barducci_metadynamics_2011,valsson_enhancing_2016,sutto_new_2012,barducci_well-tempered_2008,bussi_using_2020},
 On-the-fly probability enhanced sampling (OPES)\cite{invernizzi_rethinking_2020,invernizzi_unified_2020,invernizzi_opes_2021,invernizzi_exploration_2022}, and 
 many others. In this paper, we will focus on MetaD, 
 which relies on identifying efficient collective variables (CVs), capturing the slow modes of the process, and introducing an external bias 
potential to enhance the sampling of phase space along them. 
The ability of Metadynamics to accelerate simulations crucially
depends on the quality of the CVs\cite{invernizzi_making_2019,bussi_using_2020}. 
An optimal CV is capable of distinguishing between metastable states of interest as well as describing their interconversion dynamics. \cite{demuynck_protocol_2018,PETERS2017539}
A suboptimal CV can lead to hysteresis, and poor inference of the unbiased free 
energy surface or kinetics\cite{barducci_metadynamics_2011,besker_using_2012,invernizzi_making_2019,salvalaglio_assessing_2014,ray_rare_2022}. 

Very recently, we developed a CV-free approach for enhanced sampling, 
based on stochastic resetting (SR)\cite{blumer_stochastic_2022}. Resetting is the procedure of stopping stochastic processes, at random or fixed time intervals,
and restarting them using independent and identically distributed initial conditions. 
It has received much attention recently\cite{evans_stochastic_2020,kundu_stochastic_2022},
since it is able to expedite various processes ranging from randomized computer 
algorithms\cite{LUBY1993173,gomes_boosting_1998,montanari_optimizing_2002} and service in queuing systems\cite{bonomo_mitigating_2022}, to first-passage and search processes\cite{bressloff_queueing_2020,kusmierz_optimal_2015,bhat_stochastic_2016,chechkin_random_2018,ray_peclet_2019,robin_random_2019,evans_run_2018,pal_search_2020,bodrova_resetting_2020,luo_anomalous_2022}. 
We demonstrated the power of SR in enhanced sampling of MD simulations, 
showing it can lead to speedups of up to an order of magnitude 
in simple model systems and a molecular system\cite{blumer_stochastic_2022}. 
Moreover, we developed a method to infer the unbiased kinetics, in the absence of SR, 
from simulations with SR. 

Resetting is an appealing method due to its extreme simplicity: it can be trivially implemented in MD codes, and no CVs are required. Since finding good CVs in complex condensed phase systems is a difficult challenge, it is a potentially significant advantage. Furthermore, unlike other methods, which continuously add energy to the system, SR does not change the dynamics between resetting events. 
However, acceleration by SR is not guaranteed. A sufficient condition is that the distribution of transition times (also called first-passage times, FPTs) of the corresponding process is wide enough, i.e., it has a standard deviation that is greater than its mean\cite{pal_inspection_2022}. 
Many systems, including the models and molecular example discussed in our previous work\cite{blumer_stochastic_2022}, fulfill this criterion but there are also counter examples.

The complementary advantages and limitations of MetaD and SR raise an important question: can we combine them to obtain the best of both worlds? Since SR can be applied to any random process, we can restart MetaD simulations as previously done for unbiased simulations. This observation opens the door for using SR as a complementary tool to existing enhanced sampling procedures.

In this paper, we combine SR with MetaD for the first time. In one model system, we show that this approach leads to greater acceleration than any of them independently, even in comparison to using the optimal CV in MetaD simulations. In another model system, we restart MetaD simulations, performed with suboptimal CVs, and obtain accelerations similar to those achieved using the optimal CV. This result suggests that a straightforward application of SR can be an alternative to the
challenging task of improving a suboptimal CV. We then demonstrate this for transitions between meta-stable states of alanine tetrapeptide. Lastly, we develop a procedure to infer the unbiased kinetics from simulations combining SR and MetaD, showing an improved tradeoff between speedup and accuracy in comparison to MetaD simulations alone.

\section{Methods}

The simulations of the model systems were performed using the Large-scale Atomic/Molecular Massively Parallel Simulator\cite{LAMMPS} (LAMMPS), with MetaD introduced using PLUMED 2.7.1\cite{bonomi_plumed_2009,bonomi_promoting_2019,tribello_plumed_2014}. Alanine tetrapeptide was simulated using GROMACS 2019.6\cite{abraham_gromacs_2015} and the same PLUMED version. All simulations were carried out in the NVT ensemble at a temperature of $300 \, K$, and $10^4$ trajectories were sampled for each case.
Full simulation details are given in the SI. 

Initial velocities were sampled from the Maxwell-Boltzmann distribution, while initial positions were fixed (equivalent to sampling from a delta function positions distribution). We defined the FPT as the earliest instance in which a certain criterion was met, as specified below for each system. The COMMITTOR command in PLUMED was used for testing this criterion and stopping the simulations when it was fulfilled. For most simulations with SR, we sampled the time intervals between resetting events from an exponential distribution with a fixed resetting rate (Poisson resetting) using Python. Simulations designated for kinetics inference used constant time intervals between resetting events (Sharp resetting). If a first-passage event did not occur prior to the next resetting time, the simulation was restarted. We stress that we continue tallying the overall time until a first-passage occurred, regardless of the number of resetting events in between. 
Finally, for simulations combining SR with MetaD, we emphasize that the MetaD bias potential was zeroed after each resetting event.

\section{Results and discussion}

\subsection{A two wells model} 

\begin{figure}
  \includegraphics[width=\linewidth]{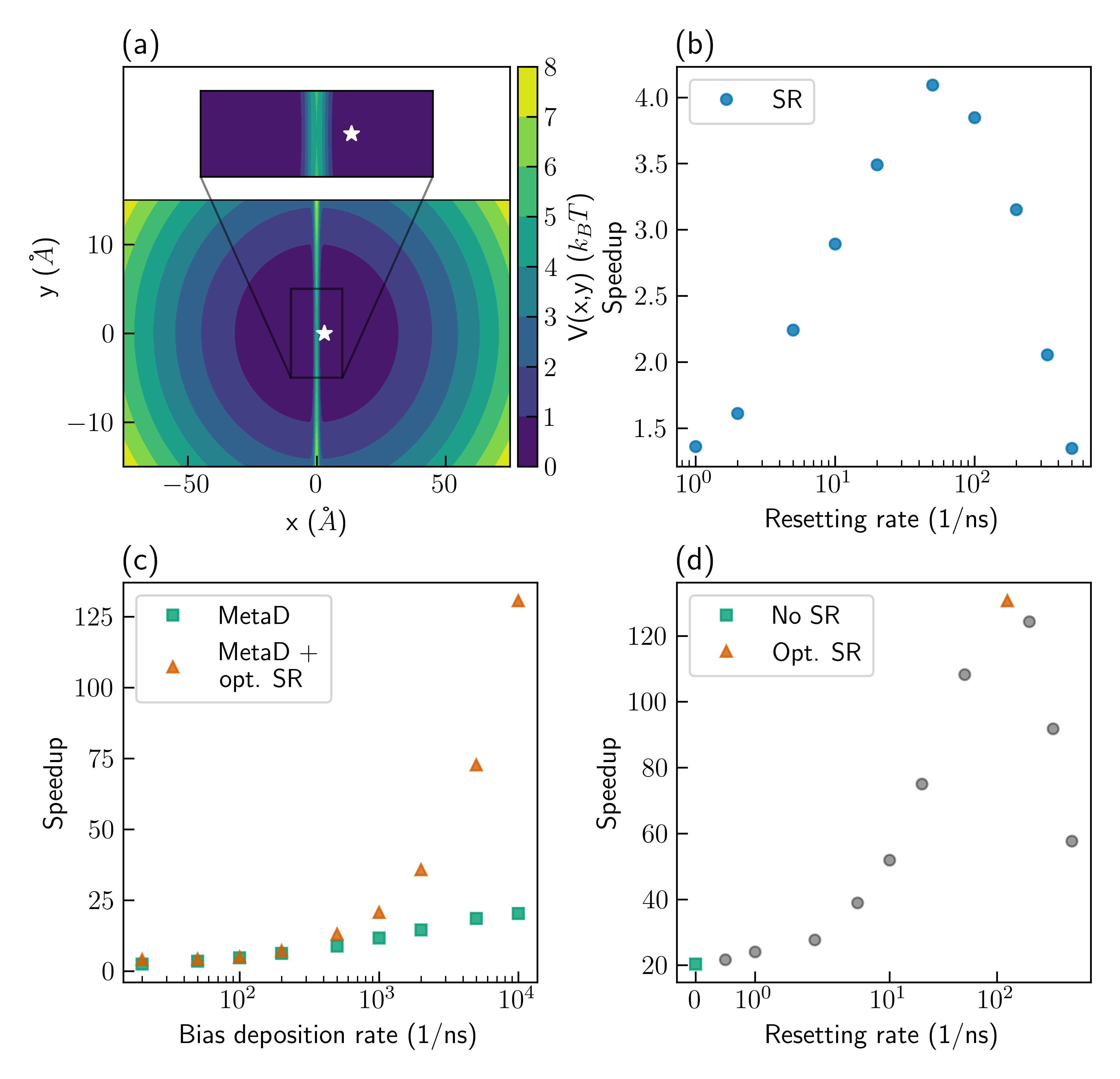}
  \caption{(a) The two wells model. The star marks the initial position. (b) Speedups obtained for SR simulations at different resetting rates. (c) Speedups obtained for MetaD simulations at different bias deposition rates (green squares), and for combined MetaD + SR simulations at an optimal resetting rate (orange triangles). (d) Speedups obtained for MetaD + SR simulations with a bias deposition rate of $10^4 \, ns^{-1}$ and different resetting rates.}
  \label{fig:FigModelA}
\end{figure}

We begin by combining SR with MetaD simulations for the first time. A simple model system is considered first, where the optimal CV is well defined.  For this model, combining SR with MetaD leads to greater speedups than MetaD independently, even when biasing with the optimal CV. The model is shown in Figure \ref{fig:FigModelA}(a). It is a two dimensional harmonic trap, divided into two states centered at $(x=\pm 3,y=0) \, \AA$ by a barrier of $5 \, k_BT$ for all y values.

The harmonic trap is soft, such that a particle in one of the wells can easily travel about $50 \, \AA$ away from the central barrier. The exact parameters of the potential are given in the SI. They were chosen such that the unbiased mean FPT (MFPT) between the wells is long ($7.5 \,ns$), but can still be sampled in unbiased simulations. The optimal CV is the x-coordinate. We follow the trajectories of a particle that was initialized at the right minimum and define the FPT criterion as arriving to the left minimum ($x \le -3 \, \AA$). 

For comparison, we first performed SR on standard MD simulations using different resetting rates. The obtained speedups are shown in Figure \ref{fig:FigModelA}(b). The speedup increases with the resetting rate, reaches a maximum value of $\sim 4$, which is obtained at a rate of $50 \, ns^{-1}$, and decreases for higher rates. This non-monotonic trend can be understood since  in the limit of high resetting rates all trajectories are restarted before a transition can occur, and the speedup drops to zero. The resetting rate which leads to maximum speedup will be referred to as the optimal resetting rate in this paper.

Next, we performed MetaD simulations without SR using the x-coordinate as a CV and varying the bias deposition rates. Other bias parameters are given in the SI. The results are shown as green squares in Figure \ref{fig:FigModelA}(c). The speedup increases with the bias rate, with a value of $\sim 20$ attained for a bias rate of $10^4 \, ns^{-1}$ (every 100 simulation steps). It is evident that MetaD leads to larger acceleration than SR for this system, giving speedups that are greater by a factor of $\sim 5$.

We then combined SR with MetaD and found that even greater speedups are
obtained (Figure \ref{fig:FigModelA}(c), orange triangles). How the combination of resetting and MetaD is done in practice is shown in Figure \ref{fig:FigModelA}(d), using the highest bias deposition rate ($10^4\,ns^{-1}$) as an example. The green square in Figure \ref{fig:FigModelA}(d) shows the speedup obtained with MetaD and no resetting.  Then, for that given bias deposition rate, we add SR at increasing rates and evaluate the resulting speedup. We again stress that the MetaD bias is zeroed at every resetting event. We observe the same qualitative behavior seen in Figure \ref{fig:FigModelA}(b), with the speedup increasing until some optimal resetting rate, highlighted with an orange triangle. We repeat this procedure for all bias rates, and present the optimal speedup by orange triangles in Figure \ref{fig:FigModelA}(c). 
Combining SR with MetaD gave additional acceleration for all bias deposition rates, with a maximal speedup of $\sim 130$ at a bias rate of $10^4\,ns^{-1}$. The corresponding optimal resetting rate was found to be $125\, ns^{-1}$, which is significantly slower than the bias deposition. The fact that SR can further accelerate MetaD simulations, even when performed with optimal CVs, is the first key result of this paper.

MetaD practitioners might wonder: 1) How to tell whether SR will accelerate my simulations?, and 2) How to identify the optimal resetting rate and what would be the resulting speedup?
Next, we show that both questions can be answered at almost no additional cost, assuming some MetaD trajectories are already available.
To answer the first question, we showed in a recent paper \cite{blumer_stochastic_2022} that a sufficient condition for acceleration of MD simulations by SR is that the ratio of the standard deviation to the MFPT (the coefficient of variation, COV) is greater than one \cite{pal_inspection_2022}. Introducing a small resetting rate is then guaranteed to lead to speedup. 
We stress that this condition holds also for resetting MetaD simulations, with the added benefit that enhanced sampling generates more transitions, and thus gives a much more reliable estimation of the COV compared to unbiased MD simulations. Moreover, if SR does not accelerate the unbiased simulations significantly, it does not mean that it will not do so for MetaD simulations, since biasing alters the FPT distribution significantly and, consequently, may also change the COV.

\begin{figure}
  \includegraphics[width=\linewidth]{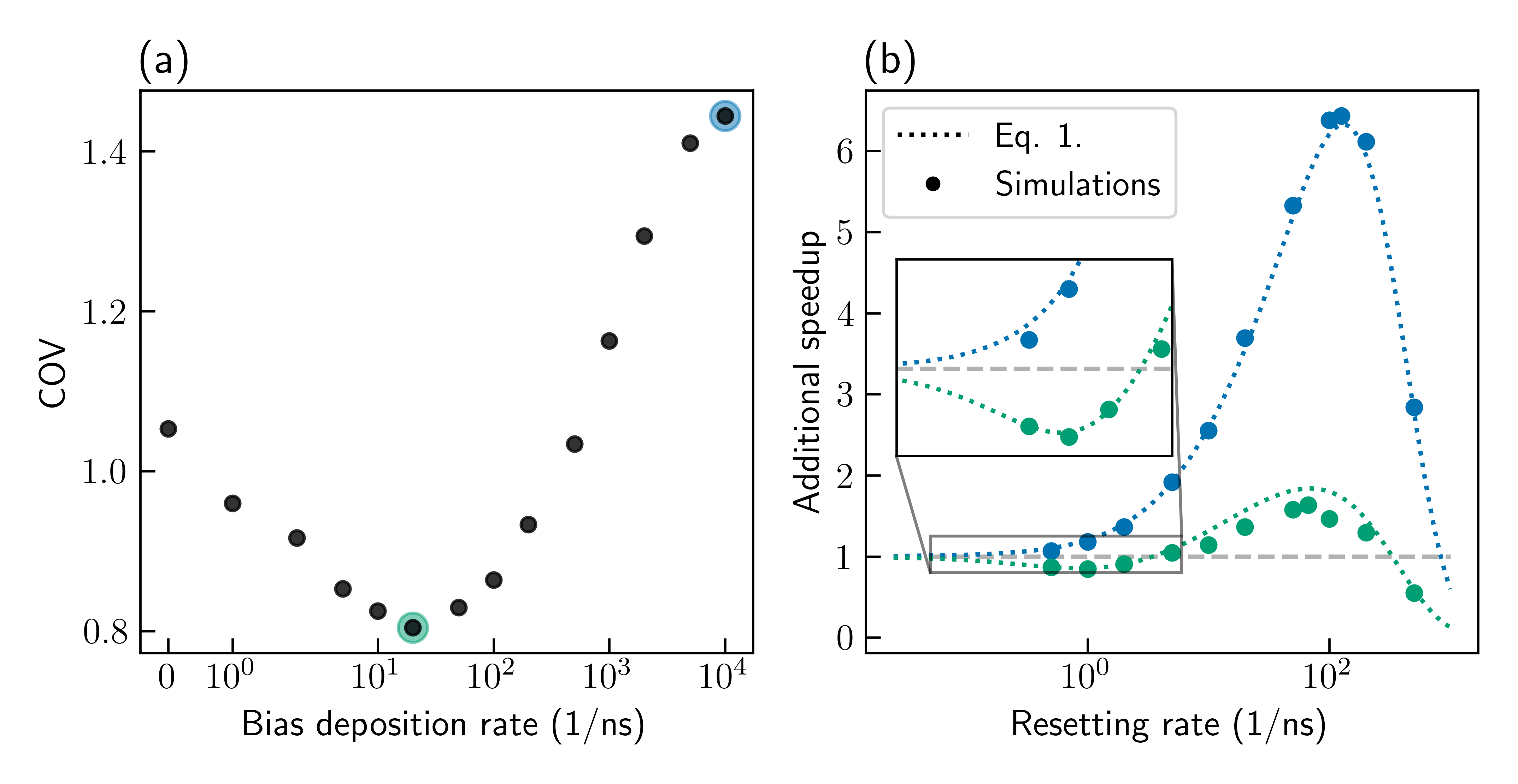}
  \caption{(a) COV of MetaD simulations with no SR. (b) Additional speedups against resetting rate for MetaD simulations with bias deposition rate of $10^4\,ns^{-1}$ (blue) or $20\,ns^{-1}$ (green), for the two wells model. Full circles present results obtained from simulations, while dotted lines present evaluations based on the FPT distribution with MetaD and no SR and using Eq. \ref{eqn:LaplaceTransformAndFPT}. The dashed gray line indicates no additional speedup.}
  \label{fig:FigExpectedSpeedups}
\end{figure}

Figure \ref{fig:FigExpectedSpeedups}(a) shows the COV of MetaD simulations (without resetting) as a function of the bias deposition rate, for the two wells model (Figure \ref{fig:FigModelA}(a)).
 The COV shows non-monotonic behavior with the bias deposition rate. It starts from a value of $1.05$ without resetting, drops to $0.80$ at an intermediate bias deposition rate and increases up to a value of $1.44$ at the highest biasing rate.
 This shows that MetaD can increase the COV significantly (leading to much higher speedups).

As for the second question, estimating the optimal resetting rate and the resulting speedup is also straightforward for MetaD simulations.
The MFPT under a resetting rate $r$ can be estimated using a simple algebraic equation \cite{reuveni_optimal_2016},
\begin{equation}
  \langle \tau \rangle_r = \frac{1-\tilde{f}(r)}{r\tilde{f}(r)}. \label{eqn:LaplaceTransformAndFPT}
\end{equation}
In Eq.~\ref{eqn:LaplaceTransformAndFPT}, $\tilde{f}(r)$ is the Laplace transform of the FPT distribution 
for the MetaD simulations, and $\langle \tau \rangle_r$ is the MFPT for MetaD simulations with SR at resetting rate $r$.
The Laplace transform is evaluated as 
\begin{equation}
    \tilde{f}(r) = \langle e^{-r \tau} \rangle \simeq \frac{1}{N} \sum_{j=1}^N e^{-r \tau_j},
\end{equation}
where $N$ is the number of MetaD trajectories, and $\tau_j$ is the FPT obtained from trajectory $j$. 
Figure \ref{fig:FigExpectedSpeedups}(b) shows the additional speedups, over MetaD without resetting, estimated using Eq.~\ref{eqn:LaplaceTransformAndFPT} (dotted lines). They are plotted as a function of the resetting rate for the two bias deposition rates highlighted with colored circles in Figure \ref{fig:FigExpectedSpeedups}(a). It is evident that the estimations match results obtained from simulations (full circles). 
While the full FPT distribution is required for an exact description 
of the behavior under SR, we previously showed\cite{blumer_stochastic_2022} that as few as a hundred 
samples are sufficient for estimating the optimal resetting rate.

Surprisingly, for intermediate bias deposition rates, some additional speedup is gained even though the COV without resetting is smaller than 1. This can be seen in Figure \ref{fig:FigExpectedSpeedups}(b), which shows an optimal speedup of $\sim 2$ for a bias deposition rate of $20 \, ns ^{-1}$, even though its COV without SR is only $0.80$.
How can it be?
A COV $<1$ indicates that introducing a small resetting rate will increase the MFPT.
In other words, the COV only indicates the initial slope of the speedup curve as a function of the resetting rate.
Interestingly, non-trivial cases where small resetting rates decelerate the process but larger ones accelerate it are also possible, as can be seen for the green curve in the inset of Figure \ref{fig:FigExpectedSpeedups}(b). For comparison, we also give the results for a bias deposition rate of $10^4 \, ns ^{-1}$, for which the COV without SR is $>1$, and the initial slope of the speedup with respect to the resetting rate is positive. The results show that, whether the value of the COV is greater or smaller than one, it is worthwhile to estimate the speedup using  Eq.~\ref{eqn:LaplaceTransformAndFPT}.

To conclude this example, we combined SR with MetaD for the first time, leading to faster acceleration than either approach separately. This is demonstrated even for a system for which the optimal CV is known. Since MetaD simulations already enhance the sampling of the underlying process, it is significantly easier to evaluate their COV than for unbiased simulations. If the COV is larger than 1, then MetaD simulations are guaranteed to be further accelerated by SR,  and Eq.~\ref{eqn:LaplaceTransformAndFPT} can be used to easily estimate by how much. If not, they may still be accelerated, which can be easily checked using Eq.~\ref{eqn:LaplaceTransformAndFPT}.

\subsection{The modified Faradjian-Elber potential}

\begin{figure}
  \includegraphics[width=0.5\linewidth]{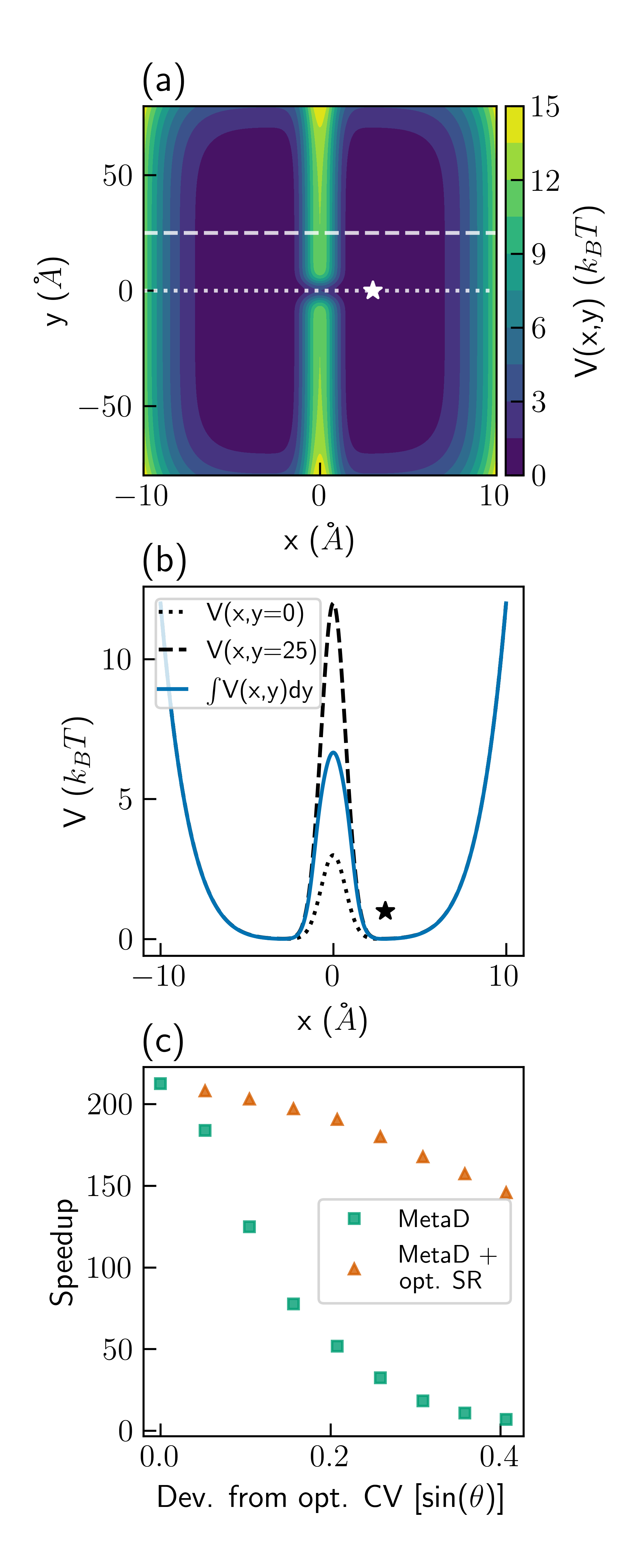}
  \caption{(a) The modified Faradjian-Elber potential. The star marks the initial position. (b) The integrated projection of the potential on the x-axis (blue), and the potential cross section at $y=0,25 \, \AA$ (dotted and dashed black lines, respectively). (c) Speedups obtained for a bias deposition rate of $10^4 \, ns^{-1}$ using suboptimal, rotated CVs, in the MetaD simulations with no SR (green squares) and with optimal SR (orange triangles). The angle between the CV and the x-axis is denoted by $\theta$.}
  \label{fig:FigModelB}
\end{figure}

As a second example, we consider a 
modified version of the two dimensional potential introduced by Faradjian and Elber when developing the milestoning enhanced sampling method~\cite{faradjian_computing_2004}. The potential is shown in Figure \ref{fig:FigModelB}(a), and full details are given in the SI. It is also composed of two symmetric wells, with minima at $(x=\pm 3,y=0) \, \AA$ that are separated by a Gaussian barrier at $x=0 \, \AA$. The barrier is higher than the first example, $12 \, k_B T$ for most $y$ values, but has a narrow saddle, only $3 \, k_B T$ high, around $y=0 \, \AA$. Figure \ref{fig:FigModelB}(b) shows slices along the x-axis at $y=0,25 \, \AA$, as well as the effective potential integrated over the entire y-axis. 

We follow the trajectories of a particle that was initialized at the right minimum and define the FPT criterion as crossing the barrier and reaching $x<-1\, \AA$. For this model, we find the same MFPT as in the two wells model. Employing SR on unbiased simulations gave an optimal speedup of $\sim 15$ at a resetting rate of $200 \, ns^{-1}$. As in the two wells model, MetaD simulations gave higher speedups than SR, with a speedup of $\sim 212$ when using the optimal CV, the x-coordinate, at a bias deposition rate of $10^4 \, ns^{-1}$. Using this optimal CV and rate, combining SR with MetaD did not lead to further acceleration of the simulations. 

However, in most real systems, the optimal CV is not known, 
and suboptimal CVs are almost always used~\cite{invernizzi_making_2019}. 
To test the efficiency of SR in such cases, we gradually reduce the quality of the CV by rotating it. The green squares in Figure \ref{fig:FigModelB}(c) show the speedup obtained as a function of the sine of the angle $\theta$ between the CV and the x-axis, which serves as a measure of the deviation from the optimal CV.  The degradation in the quality of the CV leads to a decrease in the MetaD speedup with almost no acceleration at an angle of $24^{\circ}$. However, combining SR with MetaD recovers almost all of the speedup of the optimal CV, despite the use of suboptimal CVs . This is shown by the orange triangles in Figure \ref{fig:FigModelB}(c). 
Optimizing CVs for condensed phase systems remains a difficult challenge \cite{valsson_enhancing_2016,bonati_deep_2021}. 
Our results suggest that SR may serve as an alternative, or complementary method, to 
improving CVs. Instead of using sophisticated algorithms to find better CVs \cite{bonati_deep_2021,mendels_collective_2018,piccini_metadynamics_2018,sultan_automated_2018,sidky_machine_2020,karmakar_collective_2021}, one can use SR to obtain a similar speedup at a much lower cost.
This is the second key result of this paper.

\subsection{Alanine tetrapeptide}

\begin{figure}
  \includegraphics[width=\linewidth]{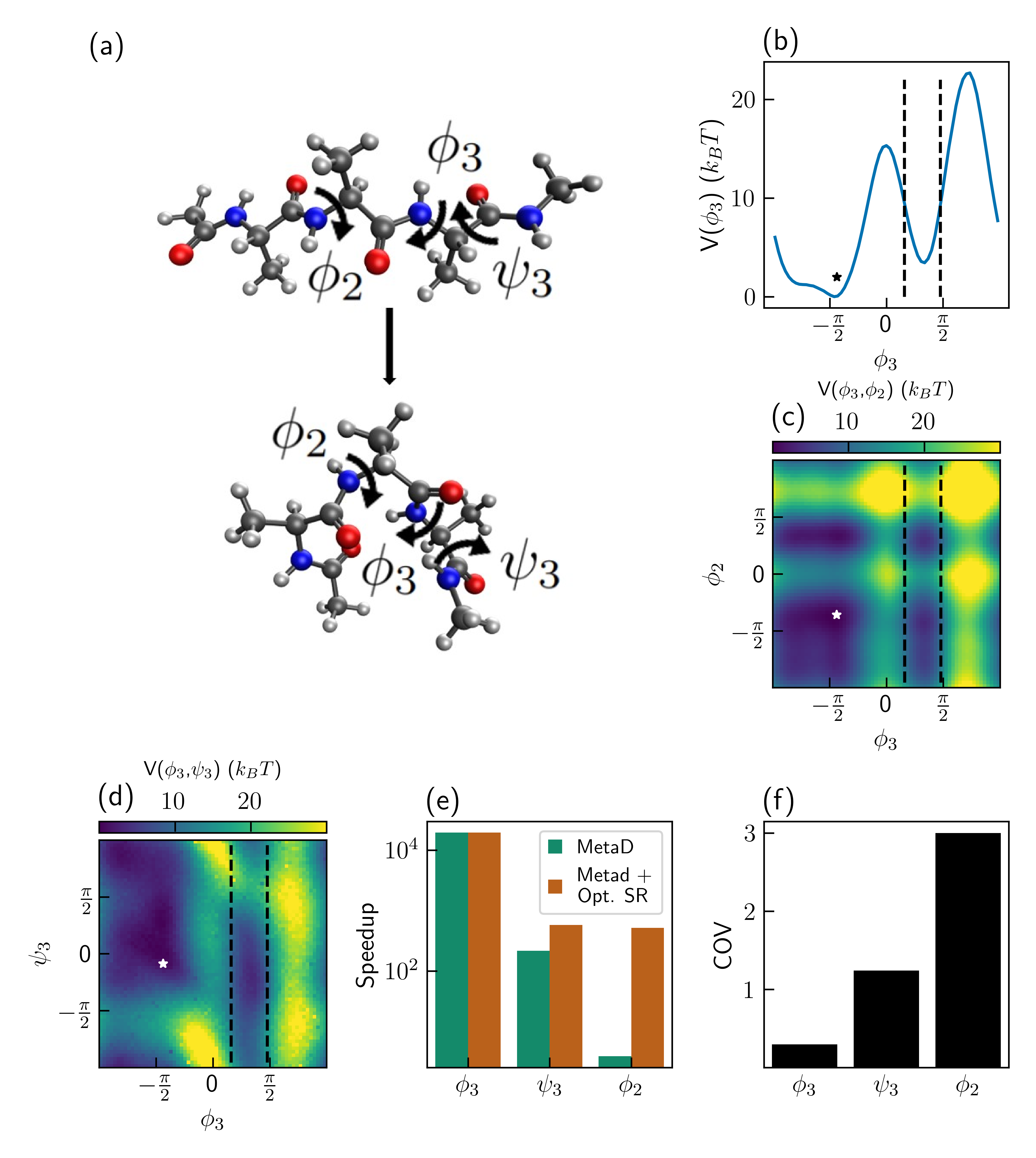}
  \caption{(a) Two conformers of alanine tetrapeptide. The white, gray, blue, and red balls represent hydrogen, carbon, nitrogen, and oxygen atoms, respectively. Free-energy surfaces of alanine tetrapeptide along (b) $\phi_3$, (c) $\phi_3, \phi_2$ and (d) $\phi_3,\psi_3$. (e) Speedup of MetaD simulations, without SR (green) and with optimal SR (orange). (f) COV of MetaD simulations without SR for different CVs.}
  \label{fig:tetra}
\end{figure}

As a final example, we demonstrate the capabilities of combining MetaD with SR on a molecular system, alanine tetrapeptide. We focus on two of its conformers: a ``folded" and ``unfolded" states, shown in Figure \ref{fig:tetra}(a). Six dihedral angles serve as important degrees 
of freedoms, with $\phi_3$ being 
the slowest one\cite{invernizzi_rethinking_2020}.
Figure \ref{fig:tetra}(b) shows the free-energy surface along $\phi_3$, 
which has two minima separated by an energy barrier of $\sim 15 \, k_BT$. 
Transitions in unbiased simulations from the unfolded state (upper configuration in panel (a), 
left basin in panel (b)) to the folded one (lower configuration in panel (a),
 right basin in panel (b)) have an estimated MFPT of $\sim 5.6\, \mu s$ (see the SI for more details).

To improve the sampling, we performed MetaD simulations using three different CVs: the angle 
$\phi_3$ serves as the optimal CV, and two adjacent angles, 
$\phi_2$ and $\psi_3$, serve as suboptimal ones. 
The two dimensional free energy surfaces as a function of all CVs
are presented in panels (c) and (d) of Figure \ref{fig:tetra}. 
They show that $\psi_3$ has some overlap and does not separate the two states as well as $\phi_3$, while there is almost no separation of the states in $\phi_2$. 
The simulations were initialized from a fixed, unfolded configuration, marked with stars in panels
(b) to (d). The first-passage criterion ($0.5 < \phi_3 < 1.5 \, rad$) is also marked in these panels, with vertical dashed lines.

Figure \ref{fig:tetra}(e) shows the speedup of MetaD simulations using different protocols, 
without SR (green) and with optimal SR (orange).
 COV values for simulations with no SR are given in panel (f). 
 As expected, using $\phi_3$ as a CV gives the greatest speedup, and a
  COV of $\sim 0.3$ for which the optimal resetting rate is $r^*=0$. Thus,  
  there is no benefit from SR in this case. 
  Suboptimal CVs show similar behavior to that observed for the Faradjian-Elber 
  potential, but for a realistic system: 
  The speedups obtained for MetaD without SR decrease when using bad CVs and the COV values increase above one.
  Namely, while MetaD simulations using $\phi_3$ as the CV gives more than four orders of magnitude speedup, simulations using $\psi_3$ and $\phi_2$ lead to accelerations by factors of only $\sim 580$ and $\sim 4$, respectively. Concurrently, the COV of $\psi_3$ is $\sim 1.24$ while for the worst CV, $\phi_2$, it is $\sim 3$.
  As a result, SR becomes more effective the poorer the CV is, giving an additional speedup of $\sim 133$ over MetaD when using $\phi_2$ as a CV. 

\subsection{Kinetics inference}

To conclude the paper, we demonstrate that SR can improve the inference of unbiased kinetics from MetaD simulations, using the case of alanine tetrapeptide as an example.
The unbiased FPT distribution can be extracted from MetaD simulations with no SR through a procedure known as infrequent MetaD
(iMetaD).\cite{tiwary_metadynamics_2013}
In this method, the MFPT is obtained by rescaling the FPT of each MetaD trajectory by an acceleration factor that depends exponentially on the external bias (see Eq. S5 in the SI). 
iMetaD assumes that no bias is deposited near the transition state, 
and that none of the basins are over-filled.
When this assumption is valid, the distribution of the rescaled FPTs matches the unbiased distribution. However, the assumption does not hold for suboptimal CVs or high bias deposition rates~\cite{salvalaglio_assessing_2014}, which result in over-deposition. 
Due to the exponential dependence of the acceleration factor on the bias, trajectories exhibiting over-deposition result in extremely large acceleration factors. They contribute unrealisticly long FPTs to the rescaled distribution, shifting the obtained MFPT from the true value by orders of magnitude. 

\begin{figure}
  \includegraphics[width=0.5\linewidth]{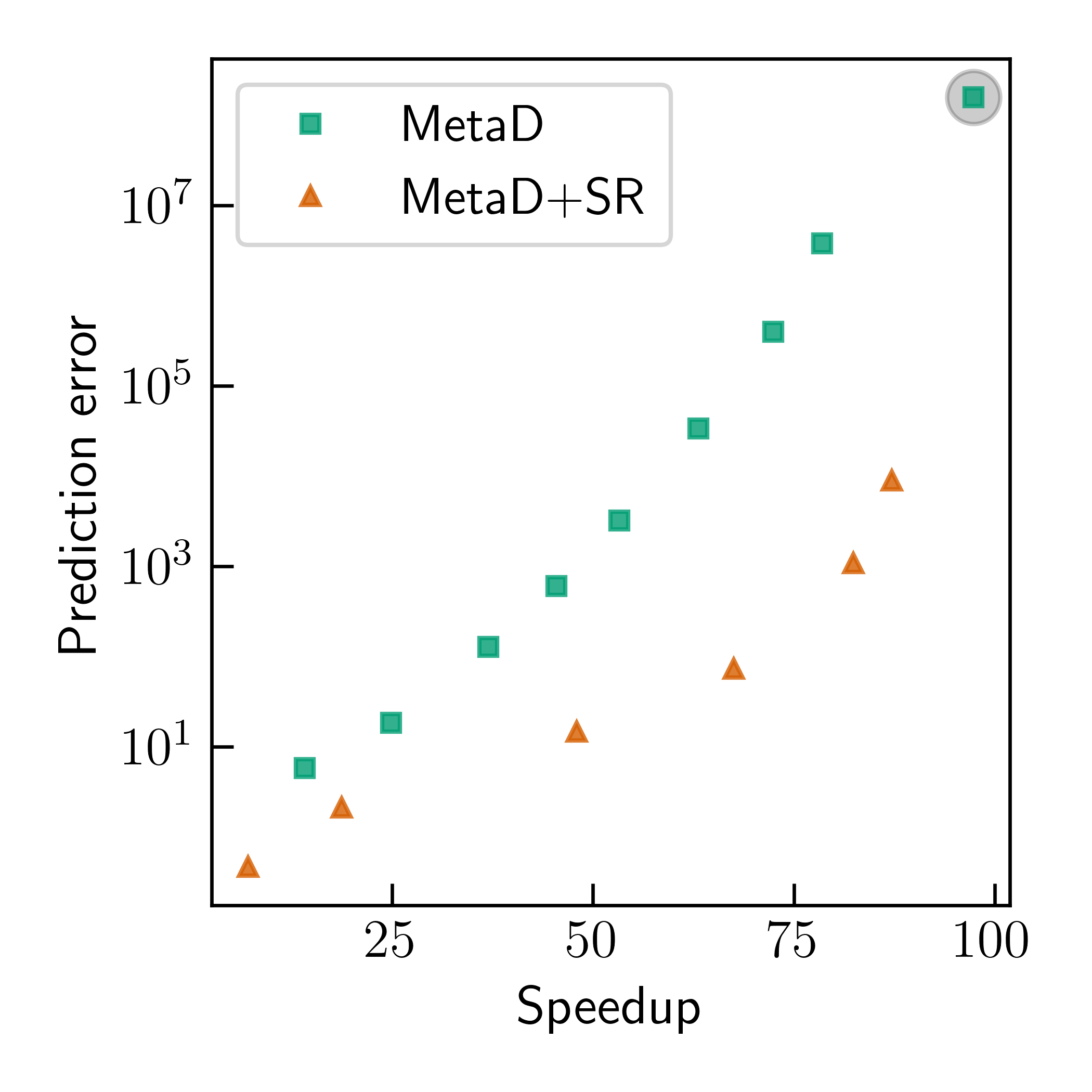}
  \caption{The error in the prediction of the unbiased MFPT as a function of speedup for iMetaD simulations at different bias deposition rates (green squares) and iMetaD simulations with SR at different resetting rates (orange triangles).} 
  \label{fig:tetraKineticsTradeOff}
\end{figure}

The inference can be improved, even with suboptimal CVs, by decreasing the bias deposition rate, resulting in a tradeoff between speedup and accuracy. This tradeoff is demonstrated in Figure \ref{fig:tetraKineticsTradeOff}. It shows (green squares) the error in the estimation of the unbiased MFPT as a function of speedup, for iMetaD simulations of alanine tetrapeptide biasing the $\psi_3$ angle, which is a suboptimal CV. The prediction error is defined as $\left| \langle \tau \rangle_{true} - \langle \tau \rangle_{est} \right| / \langle \tau \rangle_{true} $ where $\langle \tau \rangle_{true} $ is the true unbiased MFPT and $\langle \tau \rangle_{est} $ is the estimated MFPT. 

Next, we demonstrate that resetting can give a better tradeoff, reducing the error for all speedups, as shown by the orange triangles in Figure \ref{fig:tetraKineticsTradeOff}. These results were obtained by adding SR at different resetting rates to the iMetaD simulations at the highest bias deposition rate (highlighted by a circle). 

Full details explaining how to infer the unbiased MFPT from combined SR and iMetaD simulations are given in the SI. Here, we briefly provide only the key ingredients and underlying intuition. We note that between resetting events, the trajectories are standard iMetaD simulations. Moreover, due to SR, the short trajectories between restarts, can be treated as independent from one another (recall that restart also zeros previous bias). As a result, we can use the standard iMetaD rescaling procedure on each short trajectory, and then evaluate the unbiased survival probability at short times. For short enough times, even with suboptimal CVs, we will avoid over-deposition and get a good estimate of the unbiased survival. 
Finally, we assume that the survival probability decays exponentially, as commonly done in iMetaD~\cite{salvalaglio_assessing_2014}, and obtain an estimate of the unbiased MFPT from its slope. The quality of the linear fit can be used to assess the reliability of the predicted MFPT, similar to the Kolmogorov–Smirnov test used by Salvalaglio et. al in standard iMetaD simulations.\cite{salvalaglio_assessing_2014,massey_kolmogorov-smirnov_1951,noauthor_miller_nodate}

 Our results show that, for alanine tetrapetide, applying SR to iMetaD simulations and using the proposed inference procedure gives a better tradeoff than decreasing the bias deposition rate. In choosing between the two, practitioners of iMetaD might consider a simple question: given a fixed simulation time, would the error obtained be lower in a single long iMetaD trajectory or on a series of shorter iMetaD trajectories with SR? For suboptimal CVs, the acceleration factor becomes increasingly unreliable with time, due to over-deposition of the external bias. In that case, short trajectories with minimal bias are preferred, and SR may be more favorable.

\subsection{Conclusions}

We combine SR with MetaD simulations for the first time. We show that resetting can further accelerate MetaD simulations, even when the optimal CV is used. In practice, the optimal CV is almost never known and suboptimal CVs are employed.
We provide examples in which adding SR to MetaD simulations performed with poor CVs, leads to speedups comparable to using the optimal CV. 
This suggests that resetting may serve as an alternative to the challenging task of improving suboptimal CVs using sophisticated algorithms. 

Testing whether SR can accelerate simulations is very easy. Given a small number ($\sim 100$) of short MetaD trajectories, showing one transition each, we can estimate the COV and find whether SR would further accelerate the simulations, and by how much, using Eq. \ref{eqn:LaplaceTransformAndFPT}. 
Resetting can be of benefit even when it does not provide additional acceleration on top of the one attained by MetaD. We demonstrate that SR can improve the inference of the unbiased kinetics from iMetaD simulations performed with suboptimal CVs, giving a better tradeoff between speedup and accuracy for alanine tetrapeptide.

Finally, we conjecture that benefits coming from combining MetaD and SR are not limited to the examples presented herein, and are much more general. 
The reason is that MetaD, and similar methods, flatten the free energy surface.
 Previous work has shown that SR is particularly efficient for flat landscapes 
 \cite{ray_peclet_2019,ray_diffusion_2020,ray_resetting_2021}, with the extreme case being free diffusion~\cite{evans_diffusion_2011}. Thus, starting from an arbitrary free energy surface, MetaD changes it to one that is more amenable to acceleration by SR. Future method development would likely harness the power of this important observation.
  
\begin{acknowledgement}
Barak Hirshberg acknowledges support by the Israel Science Foundation (grants No. 1037/22 and 1312/22) and the Pazy Foundation of the IAEC-UPBC (grant No. 415-2023). Shlomi Reuveni acknowledges support from the Israel Science Foundation (grant No. 394/19). This project has received funding from the European Research Council (ERC) under the European Union’s Horizon 2020 research and innovation program (grant agreement No. 947731).
\end{acknowledgement}

\begin{suppinfo}
See the SI for full details of the computational setup and of the procedure to infer the unbiased kinetics from Metadynamics simulations with SR. Example input files are available on the GitHub repository~\cite{Blumer_Input_files_for}.
\end{suppinfo}

\bibliography{achemso-demo}

\end{document}

\section{General simulation details}

\subsection{Model systems}
Simulations of model potentials were performed in the Large-scale Atomic/Molecular Massively Parallel Simulator (LAMMPS) \cite{LAMMPS}. All of them were performed in the canonical (NVT) ensemble at a temperature $T=300 \, K$, using a Langevin thermostat with a friction coefficient $\gamma=0.01 \, fs^{-1}$. The integration time step was $1 \, fs$. We followed the trajectories of a single particle with mass $m=40 \, g \, mol^{-1}$, representing an argon atom. 

Metadynamics (MetaD) was implemented using PLUMED 2.7.1.\cite{bonomi_plumed_2009,bonomi_promoting_2019,tribello_plumed_2014} We used a bias factor of 10, bias height of $0.5k_BT$ and grid spacing of $0.01 \AA$. The Gaussians width was $\sigma = 1.3, 0.15 \AA$ for the two wells model and the modified Faradjian-Elber potential respectively.

\subsection{Alanine tetrapeptide}
For the simulations of alanine tetrapeptide, we used input files by Invernizzi and Parrinello\cite{invernizzi_exploration_2022}, given in PLUMED-NEST, the public repository of the PLUMED consortium\cite{Bonomi2019}, as plumID:22.003.
The simulations were performed in
GROMACS 2019.6\cite{abraham_gromacs_2015} patched with PLUMED 2.7.1\cite{bonomi_plumed_2009,bonomi_promoting_2019,tribello_plumed_2014}. They were carried out in the NVT ensemble at a temperature of $300 \, K$, using a stochastic velocity rescaling thermostat \cite{bussi_canonical_2007}, and integration time step of $2\, fs$.

We used a bias height of $0.5\,k_B T$ and grid spacing of $0.001 \, rad$ for all MetaD simulations. The bias width $\sigma$ was taken as $10\%$ of the unbiased fluctuations within the narrowest wells, $0.013,0.013,0.05 \, rad$ for angles $\phi_3,\phi_2,$ and $\psi_3$ respectively.  
In most simulations, we used a bias factor of $20$ and bias deposition rate of $5 \cdot 10^3 \, ns^{-1}$, updating the bias potential every $100$ time steps. When performing infrequent MetaD (iMetaD), we used a smaller bias factor of $10$. 

\section{Model potentials}
\label{sec:ModelPotentials}
Here, we present the exact equations and parameters of the chosen model potentials. The parameters are given such that spatial distances are in $\AA$ and potential energies are in units of $1 \, k_BT$ for a temperature of $300 \, K$.

The two wells model is described by Eq. \ref{eqn:DW}, with $A_1=\num{1e-3}$, $A_2=\num{1e-2}$, $B=5$, $C=1$.

\begin{equation}
 V(x,y) = A_1x^2 + A_2y^2 + B \exp \left(-Cx^2\right)
 \label{eqn:DW}
\end{equation}

For the modified Faradjian-Elber potential, we used Eq. \ref{eqn:FE},  with $A_1=\num{1.2e-5}$, $A_2=12$, $B=0.75$,
$\sigma_1=1$, $\sigma_2=0.5$, and $y' = 0.1y$.

\begin{equation}
V(x,y) = A_1\left(x^6+y'^6\right) + 
A_2\exp\left(-\frac{x^2}{\sigma_1^2}\right)\left[1-B\exp\left(-\frac{y'^2}{\sigma_2^2}\right)\right]
\label{eqn:FE}
\end{equation}

The modifications were made to: 1) stretch the y-axis, and 2) ensure there is a barrier also at $y=0$ by setting the value of $B \ne 1$.

\section{Estimation of mean first-passage time of transitions between conformers of alanine tetrapeptide}

To estimate the mean first-passage time (MFPT) between two conformers of 
alanine tetrapeptide, we first performed $10^4$ unbiased simulations, which were stopped when a transition was observed. However, $\sim 17.5\%$ of them did not show a transition even after $\tau_{max} = 10\, \mu s$. Therefore, unbiased simulations only provided the MFPT for transition times $\tau<\tau_{max}$, and the probability to observe a transition prior to $\tau_{max}$. We denote these quantities $\langle \tau | \tau<\tau_{max}\rangle$ and $P(\tau<\tau_{max})$ respectively. The true, unknown MFPT, $\langle \tau  \rangle$, may be written as:

\begin{equation}
\langle \tau  \rangle = P(\tau<\tau_{max})\langle \tau | \tau<\tau_{max} \rangle + 
\left(1-P(\tau<\tau_{max})\right)\langle \tau | \tau\geq\tau_{max} \rangle.
\label{eqn:probabilities}
\end{equation}

Thus, it remains to evaluate the MFPT of trajectories with $\tau\geq\tau_{max}$, denoted $\langle \tau | \tau\geq\tau_{max} \rangle$, to obtain an estimation for $\langle \tau  \rangle$. To evaluate $\langle \tau | \tau\geq\tau_{max} \rangle$, we assume that the probability density function of the process decays exponentially for times longer then some characteristic time $\tau_0$. If $\tau_0<\tau_{max}$, we can sample the rate of the decay from the tail of $P(\tau<\tau_{max})$. Practically, we plot the survival probability at times $\tau<\tau_{max}$ on a logarithmic scale and fit a linear function close to $\tau=\tau_{max}$. The slope of the fit is taken as the exponential rate $\mu$, and is used to evaluate: 

\begin{equation}
\langle \tau | \tau>=\tau_{max} \rangle = \tau_{max}+\mu^{-1}.
\end{equation}

\noindent Substituting in Eq. \ref{eqn:probabilities} yields the estimated MFPT.

About $5\%$ of MetaD simulations using the $\phi_2$ angle as CV did not show a transition. The MFPT for this case was estimated as explained above for the unbiased case. 
For this angle, we also estimated the coefficient of variation (COV), which required the standard deviation. To obtain it, for each trajectory $j$ that did not show a transition, we sampled a value $\eta_j$ from an exponential distribution with rate $\mu$ and acquired a transition time $\tau_j = \tau_{max} + \eta_j$.

\section{Kinetics inference for iMetaD simulations with SR}

The kinetics inference procedure for iMetaD simulations with SR is composed of the following steps: 
First, we obtain a set of $N$ trajectories ending in a first-passage from iMetaD simulations with SR.
We divide the full trajectories to the shorter trajectories between resetting events. These strips form a set of $M_{tot} \ge N$ shorter iMetaD trajectories, $N$ of them successfully ending with a first-passage. 
The length of each strip $t_j = n_j \Delta t$ was scaled using the standard iMetaD acceleration factor $\alpha_j$,\cite{tiwary_metadynamics_2013,valsson_enhancing_2016} to $\tilde{t}_j = \alpha_j t_j$, where
\begin{equation}
  \alpha_j = \frac{1}{n_j}\sum\limits_{i=1}^{n_j} e^{\beta V_j\left(s(t_i),t_i\right)} \label{eqn:imetad}.
\end{equation}
Here, $n_j$ is the total number of time steps in the strip, $\Delta t$ is the time step size, $V_j$ is the external bias potential, $s$ is the CV, $t_i$ is the $i$-th time step, and $\beta$ is the inverse temperature. 

Next, we evaluate the survival function, defined as $S\left(\tilde{t}\right) =M_{\tilde{t_j}>\tilde{t}}/M_{tot}$, where $M_{\tilde{t}_j>\tilde{t}}$ being the number of strips with rescaled length that is larger than $\tilde{t}$. 
Finally, we assume that the underlying true FPT distribution is exponential. For an exponential FPT distribution, $\log\left(S\left(\tilde{t}\right)\right)$ decays linearly with a slope of $-\langle \tau \rangle^{-1}$, where $\langle \tau \rangle$ is the MFPT. We perform a linear fit to the obtained survival function, and use its slope to estimate the unbiased MFPT.
To include all trajectories in the analysis, we took $\tilde t$ to be smaller than the $\tilde {t}_j$ of the shortest trajectory that did not show a transition. 

\begin{figure}
  \includegraphics[width=\linewidth]{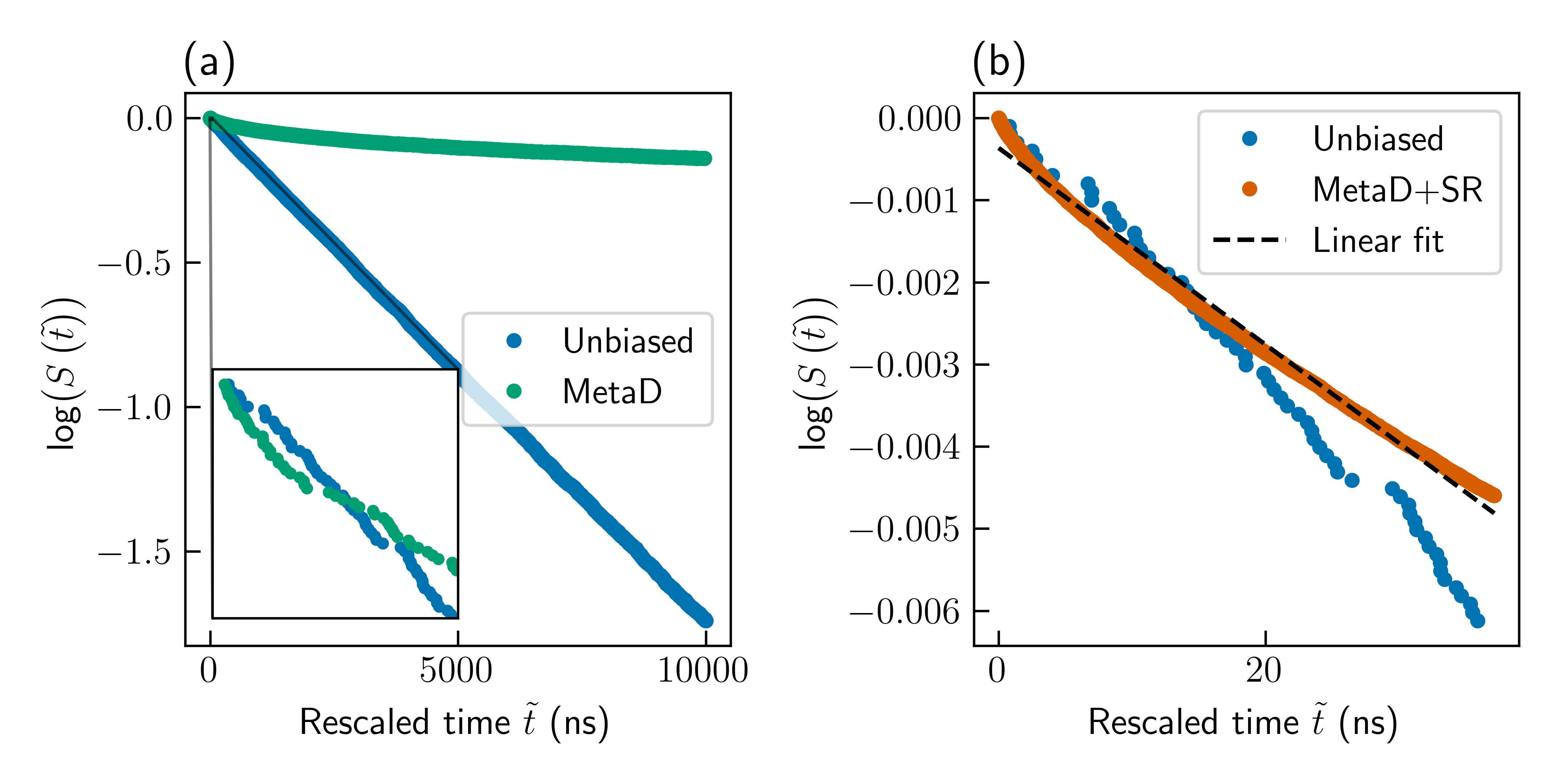}
  \caption{a) The survival function for unbiased trajectories (blue) and MetaD trajectories (green). The inset shows a zoom in on short-times. b) The survival function for MetaD trajectories with SR (orange). A linear fit is shown in black.}
  \label{fig:survivalFunction}
\end{figure}

\begin{figure}
  \includegraphics[width=0.5\linewidth]{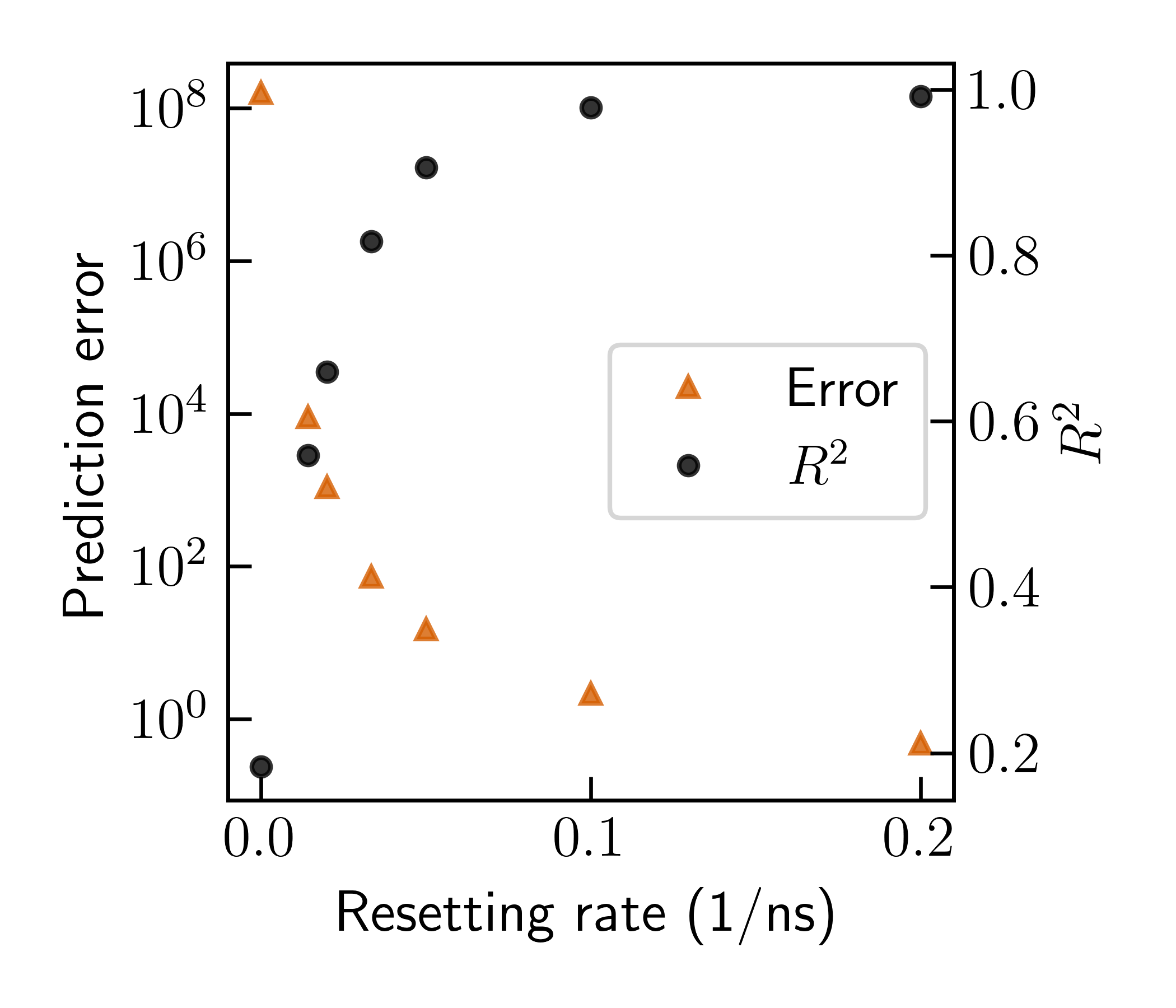}
  \caption{Prediction error (orange) and $R^2$ values as a function of resetting rate.}
  \label{fig:assessReliability}
\end{figure}

This procedure is demonstrated for alanine tetrapeptide.
Figure \ref{fig:survivalFunction}(a) shows $\log\left(S\left(\tilde{t}\right)\right)$ for iMetaD simulations and no SR using the sub-optimal CV $\psi_3$ (in green). The true survival function, as obtained from unbiased simulations, is given in blue. Due to bias over-deposition, the resulting survival function decays much slower than the true one. However, at short times, where the bias is minimal, even a sub-optimal CV gives a survival function that tracks the unbiased one.
The benefit of resetting is in providing excessive sampling of the short time region, leading to more reliable estimation of the exponential decay of the survival. The survival function estimated from iMetaD with SR is given in panel (b), showing an improved agreement with the unbiased results. The quality of the linear fit (black line) provides an assessment for the reliability of the results. Specifically, we use the Pearson correlation coefficient $R$ between the samples and the linear fit. Figure \ref{fig:assessReliability} shows $R^2$ next to the associated prediction error, as a function of resetting rate. It reaches $R^2\to1$ for high resetting rates, leading to minimal error.

\bibliography{si}